\documentclass[pra,twocolumn,amssymb,amsmath,nobibnotes,aps,showpacs]{revtex4-1}

\usepackage{graphics}
\usepackage{graphicx}
\usepackage{braket}
\usepackage{float}
\usepackage{multirow}
\usepackage{color}

\begin{document}

\title{Collision cross sections for the thermalization of cold gases}

\author{Matthew D. Frye and Jeremy M. Hutson}

\affiliation{Joint Quantum Centre (JQC) Durham/Newcastle, Department of
Chemistry, Durham University, South Road, Durham DH1 3LE, United Kingdom}

\date{\today}

\begin{abstract}
The collision cross section that controls thermalization of gas mixtures is the
transport cross section $\sigma_\eta^{(1)}$ and not the elastic cross section
$\sigma_{\rm el}$. The two are the same for pure s-wave scattering but not when
higher partial waves contribute. We investigate the differences between them
for prototype atomic mixtures and show that the distinction is important at
energies above 100 $\mu$K for LiYb and 3 $\mu$K for RbYb and RbCs. For simple
systems both $\sigma_\eta^{(1)}$ and $\sigma_{\rm el}$ follow universal energy
dependence that depends only on the s-wave scattering length when expressed in
reduced length and energy units.
\end{abstract}


\maketitle

\section{Introduction}

The scattering length for interaction between a pair of atoms or molecules is a
key quantity in ultracold physics. It determines the cross sections for
ultracold collisions and the energy of a Bose-Einstein condensate. Manipulating
the scattering length with applied magnetic, electric or optical fields
provides the main way to control ultracold gases, allowing the investigation of
condensate collapse, solitons, molecule formation and many other phenomena.

Precise determinations of scattering lengths may be achieved by fitting the
energies of high-lying bound states or the positions of zero-energy Feshbach
resonances as a function of applied field. However, in the early stages of
investigating a new ultracold mixture, approximate scattering lengths are often
obtained from experimental studies of interspecies thermalization rates, which
in turn depend on collision cross sections.

It is often supposed that the rate of thermalization is determined by the
elastic cross section $\sigma_{\rm el}$
\cite{deCarvalho:1999,Delannoy:2001,Tokunaga:2011}, which is related to the
differential cross section $d\sigma/d\omega$ by
\begin{equation}
\sigma_{\rm el} = \int \frac{d\sigma}{d\omega} d\omega
= \int \frac{d\sigma}{d\omega} \sin\Theta\,d\Theta,
\end{equation}
where $\Theta$ is the deflection angle in the center-of-mass frame. However,
collisions that cause only small deflections of the collision partners
contribute fully to the elastic cross section but make very little contribution
to kinetic energy transfer and thus to thermalization. The appropriate cross
section that takes this into account is the transport cross section
$\sigma_\eta^{(1)}$,
\begin{equation}
\sigma_\eta^{(1)} = \int \frac{d\sigma}{d\omega} (1-\cos\Theta) \sin\Theta\,d\Theta,
\end{equation}
which has been used extensively in the context of transport properties at
higher temperatures \cite{Maitland:1981,Liu:1979}. It determines the binary
diffusion coefficient for a mixture, and contributes to the shear viscosity
coefficient.

The relevance of $\sigma_\eta^{(1)}$ to thermalization of ultracold gases has
been pointed out by Anderlini and Gu\'ery-Odelin \cite{Anderlini:2006} (who
call it $\widetilde{\sigma}$), but no study of its behavior has been made for
the conditions relevant to thermalization of ultracold atoms and molecules. The
purpose of the present paper is to explore the behavior of $\sigma_\eta^{(1)}$
and to compare it with $\sigma_{\rm el}$ for cold and ultracold collisions. For
this purpose we will consider two topical systems, LiYb and RbYb, for both of
which there have been recent studies of thermalization aimed at determining
scattering lengths \cite{Ivanov:2011, Hara:2011, Baumer:thesis:2010,
Baumer:2011}. $\sigma_{\rm el}$ and $\sigma_\eta^{(1)}$ are equivalent when
$d\sigma/d\omega$ is isotropic, which is true both for classical hard-sphere
collisions and for quantum scattering at limitingly low energy (in the s-wave
regime). However, we will show that there are significant differences between
$\sigma_{\rm el}$ and $\sigma_\eta^{(1)}$ for realistic potentials, and that
these should be taken into account when using thermalization results to
estimate scattering lengths, particularly in the energy regime where s-wave and
p-wave collisions make comparable contributions. In addition, we will show
that, for systems of this type, the scattering properties of low-$L$ partial
waves with $L>0$ are almost universal functions of the s-wave scattering length
$a_{\rm s}$, and that the behavior of both $\sigma_{\rm el}$ and
$\sigma_\eta^{(1)}$ in the few-partial-wave regime can be predicted from a
knowledge of $a_{\rm s}$ alone.

Expansion of the differential cross section allows an alternative expression
for $\sigma_\eta^{(1)}$ to be written in terms of partial-wave phase shifts
$\delta_L$ \cite{Anderlini:2006},
\begin{equation}
\sigma_\eta^{(1)}=\frac{2\pi}{k^2}\sum_{0 \leq L \leq L' < \infty} \alpha_{L,L'}
\sin \delta_L \sin \delta_{L'} \cos(\delta_L - \delta_{L'})
\label{sig_eta_alt}
\end{equation}
where $E=\hbar^2k^2/2\mu$ is the collision energy, $\mu$ is the reduced mass,
and
$$\alpha_{L,L'}=(2-\delta_{L,L'})(2L+1)(2L'+1)\int_{-1}^1(1-x)P_L(x)P_{L'}(x),$$
which evaluates to $\alpha_{L,L} = 4L+2$, $\alpha_{L,L+1}=-(4L+4)$, $\alpha=0$
otherwise. The equivalent expression for $\sigma_{\rm el}$ contains only the
terms with $L=L'$, so the difference between the two cross sections takes the
form of a set of interference terms between partial waves with $\Delta L=\pm1$,
which may be either positive or negative.

\begin{figure*}[tbh]
 \includegraphics[width=170mm]{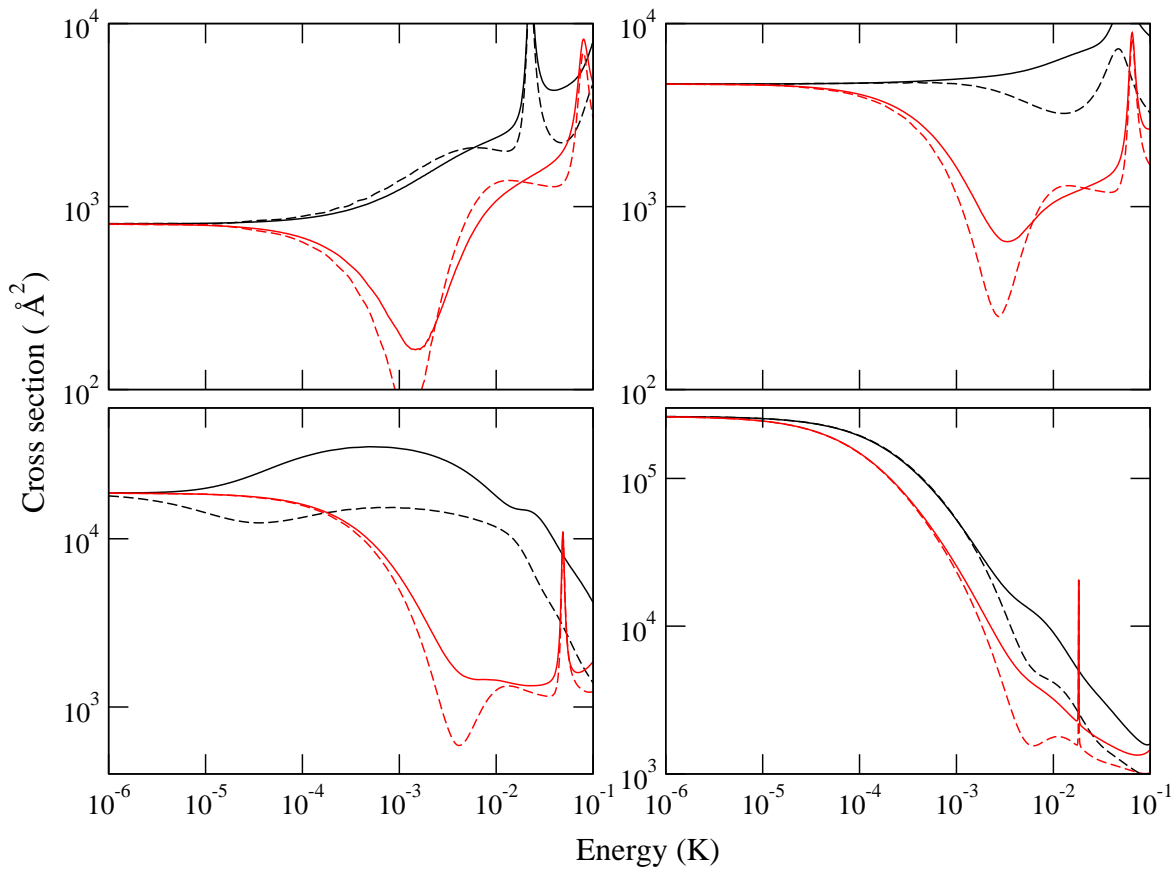}
\caption{LiYb cross sections. $\sigma_{\rm el}$ (solid lines) and
$\sigma_\eta^{(1)}$ (dashed lines) for positive (black) and negative (red)
signs of the scattering length for different values of the magnitude of the
scattering length: (a) $|a_{\rm s}|=8$~\AA\ (b) $|a_{\rm s}|=\bar{a}=19.3$~\AA\ (c)
$|a_{\rm s}|=2\bar{a}=38.6$~\AA\ (d) $|a_{\rm s}|=7.5\bar{a}=145$~\AA\
 \label{fig:LiYb}}
\end{figure*}

\section{Numerical results}

Systems such as LiYb and RbYb, made up an alkali-metal atom ($^2$S) and a
closed-shell atom ($^1$S), exhibit very narrow Feshbach resonances due to
coupling between the alkali-metal hyperfine states due to the dependence of the
hyperfine coupling on the internuclear distance $R$ \cite{Zuchowski:RbSr:2010,
Brue:LiYb:2012, Brue:AlkYb:2013}. However, these resonances have widths of 100
mG or less; collisions far from resonance can be accurately described by
single-channel calculations that neglect both electron and nuclear spin and are
independent of magnetic field. In the present work we solve the single-channel
Schr\"odinger equation using the MOLSCAT package \cite{molscat:v14}. The SBE
post-processor \cite{SBE} is then used to calculate $\sigma_\eta^{(1)}$ from
S-matrix elements as described by Liu {\em et al.}\ \cite{Liu:1979}. We use
interaction potentials for LiYb \cite{Zhang:2010} and RbYb
\cite{Brue:AlkYb:2013} from electronic structure calculations, with a fixed
long-range $C_6$ coefficient and the short-range potential scaled by a factor
$\lambda$ ($0.944<\lambda<1.033$) to adjust the s-wave scattering lengths as
required.

For $^6$Li$^{174}$Yb, recent thermalization experiments suggest an s-wave
scattering length $|a_{\rm s}|\approx8$~\AA\ \cite{Ivanov:2011, Hara:2011}, but
cannot determine the sign in the low-temperature regime investigated, where
only s-wave scattering contributes. However, the sign could be determined from
thermalization measurements at higher energies, where higher partial waves
contribute. Figure \ref{fig:LiYb}(a) shows $\sigma_{\rm el}$ and
$\sigma_\eta^{(1)}$ for $a_{\rm s}=\pm8$~\AA: it may be seen that the cross
sections for positive and negative scattering lengths deviate from one another
substantially above 40 $\mu$K, and $\sigma_{\rm el}$ and $\sigma_\eta^{(1)}$
start to differ significantly in the same region. Thus measurements at
temperatures high enough to determine the sign of the scattering length should
take into account the difference between $\sigma_{\rm el}$ and
$\sigma_\eta^{(1)}$.

The remaining panels of Fig.\ \ref{fig:LiYb} show analogous results for other
values of $|a_{\rm s}|$, in order to illustrate the range of possible behaviour
for other systems. These are chosen to be multiples of the mean scattering
length $\bar a$ \cite{Gribakin:1993}, which is 19.3~\AA\ for $^6$Li$^{174}$Yb.
The corresponding energy scale is $\bar{E}=\hbar^2/2\mu\bar{a}^2$, which is
11.2~mK here. It may be seen that in most cases $\sigma_{\rm el}$ and
$\sigma_\eta^{(1)}$ are reasonably similar at energies up to about 100 $\mu$K
(about $10^{-2}\bar{E}$); this may be compared with the p-wave barrier height
of 2.8 mK. However, the difference between $\sigma_{\rm el}$ and
$\sigma_\eta^{(1)}$ begins at much lower energies (near 1~$\mu$K) for values of
$a_{\rm s}$ near $+2\bar a$. This occurs because angular-momentum-insensitive
quantum defect theory (AQDT) predicts a p-wave shape resonance close to zero
collision energy when $a_{\rm s}=+2\bar a$, for a potential curve that behaves
as $-C_6 R^{-6}$ at long range \cite{Gao:2000}. The resonance-enhanced p-wave
scattering introduces interference terms into Eq.\ \ref{sig_eta_alt} even at
very low energy.

AQDT predicts that, in the absence of Feshbach resonances, low-energy elastic
scattering for all partial waves can be described by a single parameter which
is uniquely linked to the ratio $a_{\rm s}/\bar{a}$. Hence, any two systems
which have the same $a_{\rm s}/\bar{a}$ should have identical scattering
properties in suitably reduced units within a certain energy range around
threshold. Full details are given by Gao \cite{Gao:2001, Gao:2009}.

\begin{figure}[tbh]
 \includegraphics[width=\linewidth]{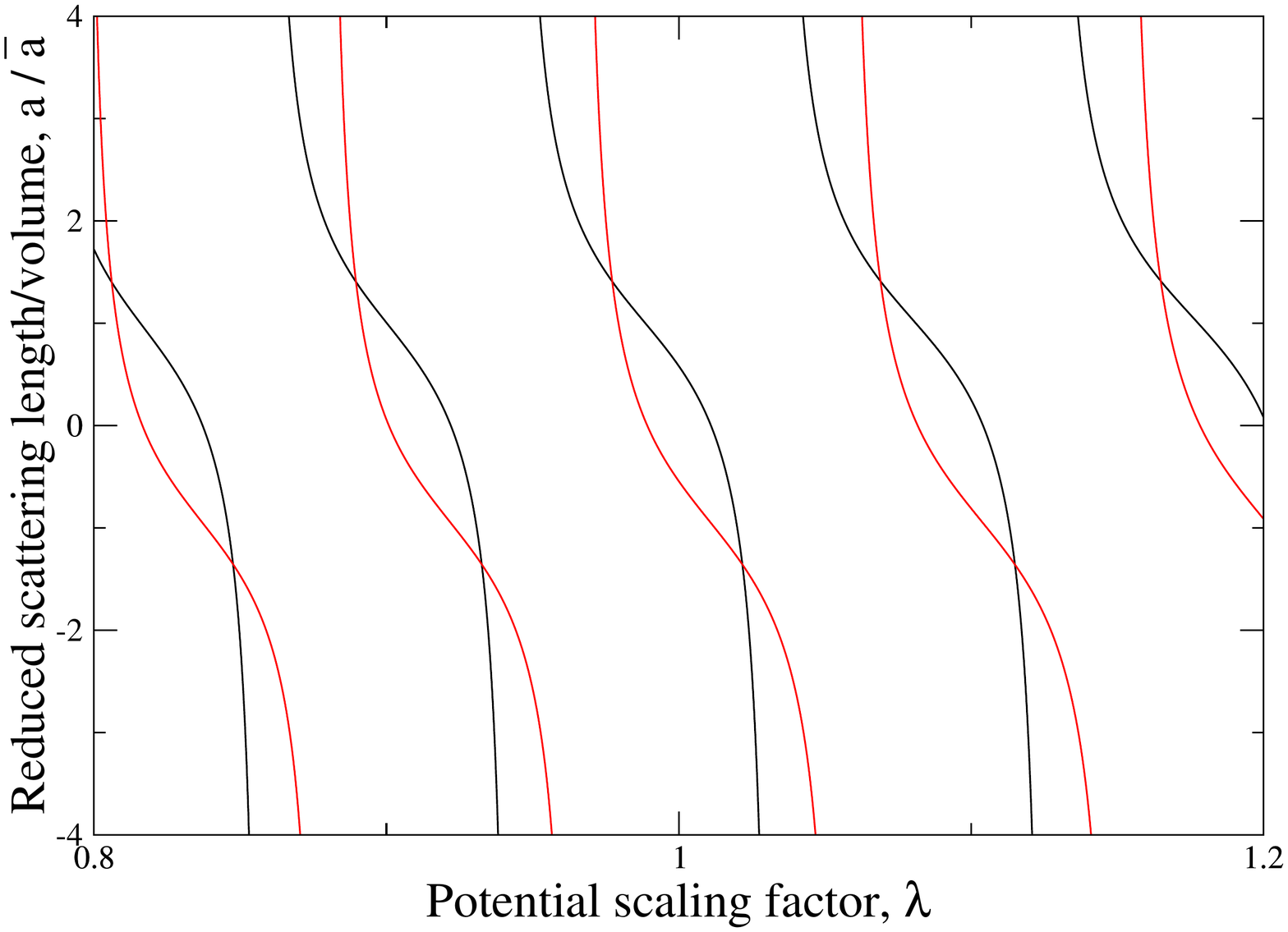}
 \includegraphics[width=\linewidth]{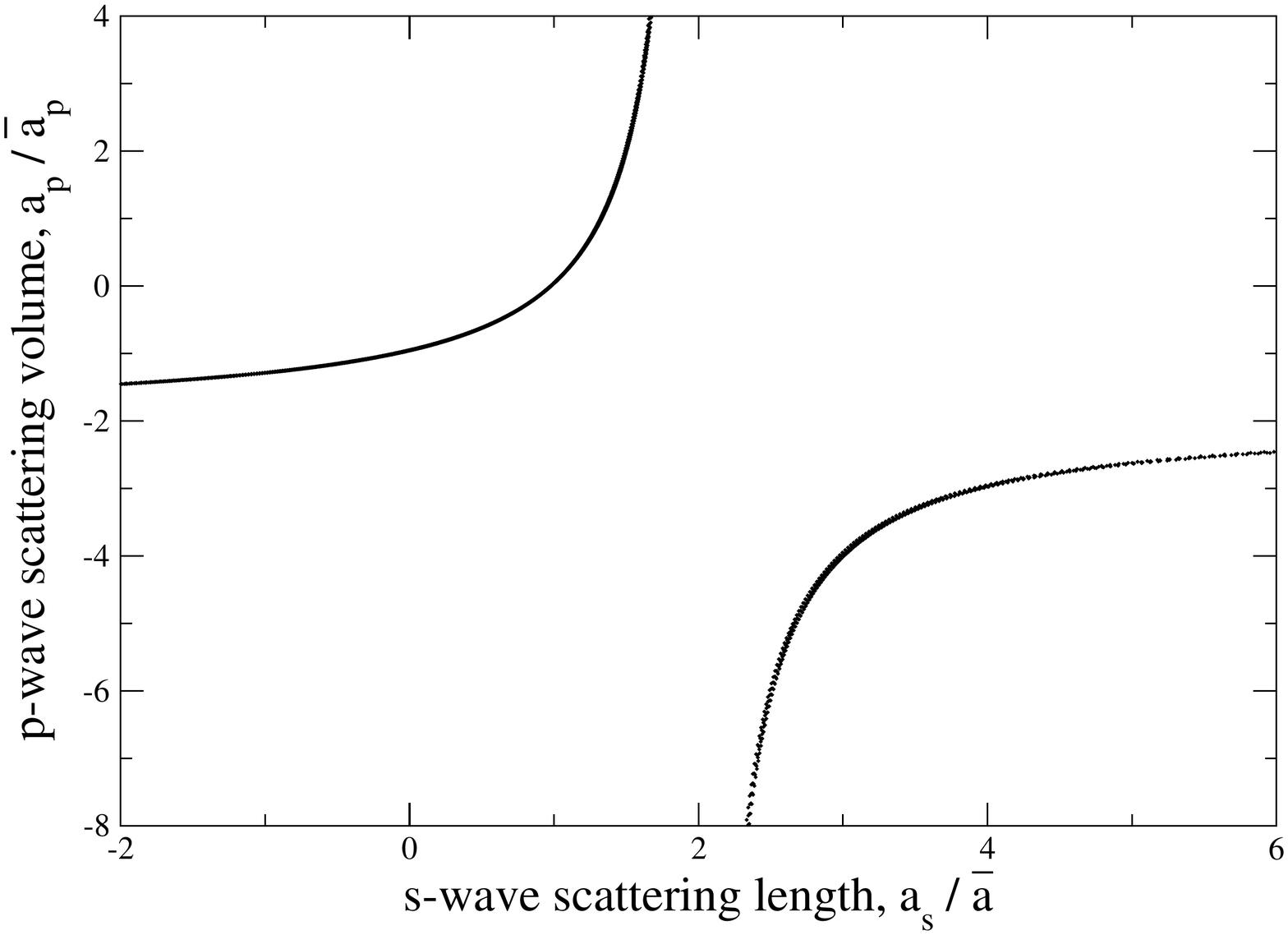}
\caption{Comparison of s-wave and p-wave scattering lengths/volumes, in units of
$\bar{a}$, across a wide range of the potential scaling factor, $\lambda$.
 \label{fig:a_s_p}}
\end{figure}

The relationship between scattering in different partial waves is conveniently
demonstrated by considering the relationship between the s-wave scattering
length and the equivalent quantities for higher partial waves \cite{Gao:2009}
(which are no longer lengths but volumes or hypervolumes). For example the
p-wave scattering volume $a_{\rm p}$ is predicted by AQDT to be
\begin{equation}
\frac{a_{\rm p}}{\bar{a}_{\rm p}}=-2\left[1+\frac{1}{a_{\rm s}/\bar{a} -2}\right].
\label{eq:apas}
\end{equation}
where $\bar{a}_{\rm p}$ is the mean p-wave scattering volume \cite{Gao:2009}.
Fig.\ \ref{fig:a_s_p}(a) shows $a_{\rm s}$ and $a_{\rm p}$ for LiYb as a
function of a potential scaling factor, $\lambda$, as it is adjusted between
0.8 and 1.2, while Fig.\ \ref{fig:a_s_p}(b) shows $a_{\rm p}$ as a function of
$a_{\rm s}$ over the same range of $\lambda$. It may be seen that $a_{\rm p}$
is indeed a nearly single-valued function of $a_{\rm s}$, as predicted by Eq.\
\ref{eq:apas}.

\begin{figure}[tbh]
 \includegraphics[width=\linewidth]{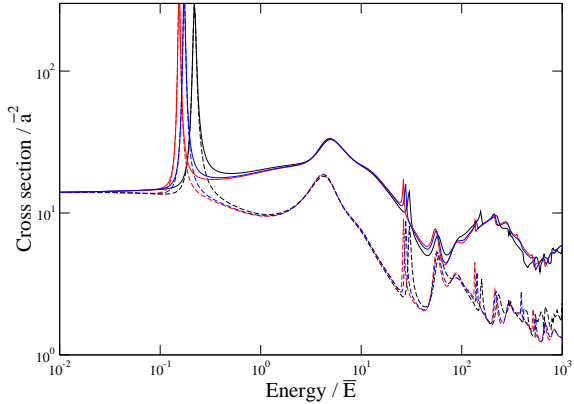}
\caption{Comparison of $\sigma_{\rm el}$ (solid lines) and $\sigma_\eta^{(1)}$
(dashed lines) in reduced units with a scattering length of $a_{\rm
s}=1.05\bar{a}$ for LiYb (red) and RbYb (blue) (mostly indistinguishable on
this scale), compared with analytic AQDT (black) results. The length and energy
scaling factors $\bar{a}$ and $\bar{E}$ are 19.3~\AA\ and 11.2~mK for LiYb and
39.6~\AA\ and 270~$\mu$K for RbYb.} \label{fig:compare}
\end{figure}

To test the extent of the universal relationship, we have carried out
calculations of $\sigma_{\rm el}$ and $\sigma_\eta^{(1)}$ for RbYb and LiYb for
potentials scaled to give identical values of $a_{\rm s}/\bar{a}$. The results
in reduced units are compared for the case of $a_{\rm s}=1.05\bar{a}$ in Fig.\
\ref{fig:compare}. According to AQDT, values of $a_{\rm s}$ slightly greater
than $\bar{a}$ produce a d-wave shape resonance at low energy, and this appears
as a prominent feature for both species in Fig.\ \ref{fig:compare}. It may be
seen that deviations between $\sigma_{\rm el}$ and $\sigma_\eta^{(1)}$ are
again significant at collision energies above about $10^{-2}\bar{E}$. However,
apart from small differences in resonance positions due to the effects of
potential terms other than $-C_6R^{-6}$, the results in reduced units are
remarkably similar for LiYb and RbYb up to energies around $400\bar{E}$, which
is about 4 K for LiYb and 100 mK for RbYb. Similar agreement was obtained for
other values of the scattering length. The calculation on full potential curves
may also be compared with those of pure AQDT \cite{Gao:2001,
Gao:AQDTroutines}, shown in black in Fig.\ \ref{fig:compare}.

\begin{figure}[tbh]
 \includegraphics[width=\linewidth]{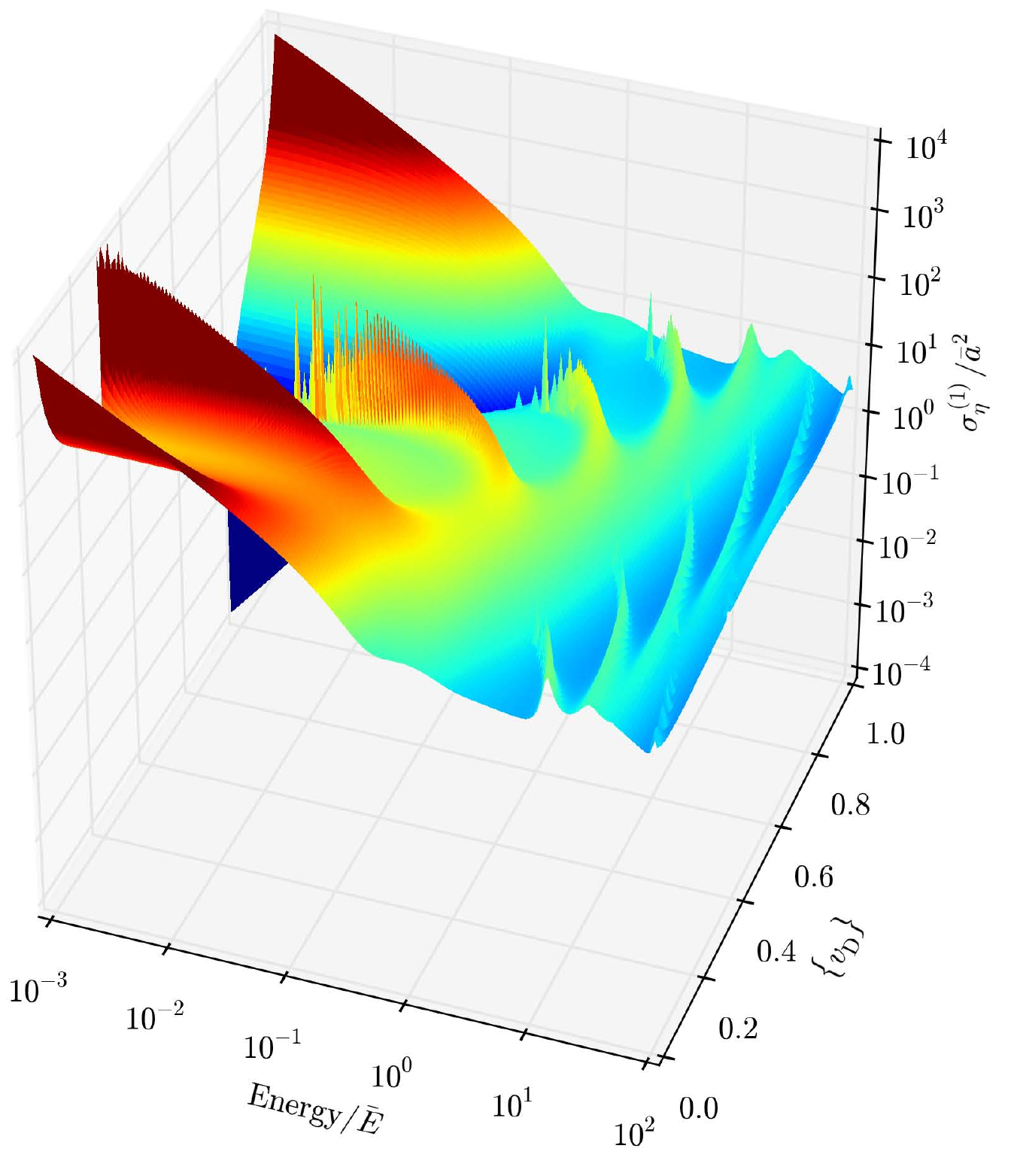}
\caption{The cross section $\sigma_\eta^{(1)}$ for RbYb, as a function of the
fractional part $\{v_{\rm D}\}$ of the quantum number at dissociation and the
energy in reduced units. The energy scaling factor $\bar{E}$ is 270~$\mu$K for
RbYb. The spikes visible at the left-hand end of some narrow ridges are
artefacts of the finite grid used for plotting.} \label{fig:surface}
\end{figure}

The universality shown in Fig.\ \ref{fig:compare} allows us to discuss RbYb in
terms of the LiYb results shown in Fig.\ \ref{fig:LiYb}, with appropriate
scaling of energies and cross sections. The scattering lengths for RbYb vary
substantially with Rb and Yb isotope \cite{Baumer:thesis:2010, Baumer:2011,
Muenchow:2011, Muenchow:thesis:2012, Brue:AlkYb:2013, Borkowski:2013}. For
$^{87}$RbYb they range from a very small value for $^{87}$Rb$^{170}$Yb to a
very large value for $^{87}$Rb$^{174}$Yb, but there are no Yb isotopes that
have $a_{\rm s}$ values near $\bar{a}$.

Fig.\ \ref{fig:surface} shows $\sigma_\eta^{(1)}$ for RbYb as a function of the
fractional part $\{v_{\rm D}\}$ of the quantum number at dissociation $v_{\rm
D}$ and the energy in reduced units. Different values of $v_{\rm D}$ were
obtained by scaling the potential of ref.\ \cite{Brue:AlkYb:2013} as described
above for LiYb, but could equivalently have been achieved by scaling the
reduced mass. The s-wave scattering length is related to $v_{\rm D}$ by
\begin{equation}
a_{\rm s} = \bar{a}\left[1-\tan\left(\frac{\pi}{4}\right)
\tan\left(v_{\rm D}+\textstyle\frac{1}{2}\right)\pi\right].
\end{equation}
It thus has a pole whenever $v_{\rm D}$ is integer and is large and positive
when $\{v_{\rm D}\}$ is small. Fig.\ \ref{fig:surface} thus shows large peaks
when $\{v_{\rm D}\}$ is 0 or 1 and a trough when $\{v_{\rm D}\}=\frac{3}{4}$,
so that $a_{\rm s}=0$. In addition, there are strong features due to shape
resonances, which sharpen and eventually become invisible as the energy
decreases. The ridge that points towards $\{v_{\rm D}\}=\frac{1}{4}$, $a_{\rm
s}=2\bar{a}$ is due to a p-wave resonance, while the ones that points towards
$\{v_{\rm D}\}=\frac{1}{2}$, $a_{\rm s}=\bar{a}$ and $\{v_{\rm
D}\}=\frac{3}{4}$, $a_{\rm s}=0$ are due to d-wave and f-wave resonances,
respectively. A series of ridges due to shape resonances with higher partial
waves may also be seen at higher energies, and can be followed at least up to
$L=9$. Their positions closely follow the prediction of AQDT, which is that, at
zero energy, resonances with $L\ge4$ occur at the same location as those with
$L-4$. Fig.\ \ref{fig:surface} would look very similar for any other
single-channel system with potential of the form $-C_6 R^{-6}$ at long range.

The situation is somewhat more complicated for pairs of alkali-metal atoms and
other systems with extensive Feshbach resonances. The overall magnitude of the
differences between $\sigma_{\rm el}$ and $\sigma_\eta^{(1)}$ are likely to be
similar in such systems. AQDT still applies usefully to the {\em background}
scattering (away from Feshbach resonances), and in such regions the
``universal" behavior of $\sigma_{\rm el}$ and $\sigma_\eta^{(1)}$ will still
apply, at least at relatively low energies. However, understanding the detailed
behaviour, including resonant effects, requires coupled-channel calculations
using accurate potential curves.

\begin{figure}[tbh]
 \includegraphics[width=\linewidth]{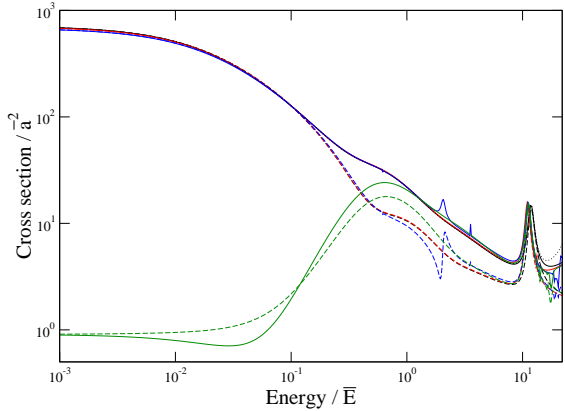}
\caption{Coupled-channel calculations of $\sigma_{\rm el}$ (solid lines) and
$\sigma_\eta^{(1)}$ (dashed lines) for RbCs at various magnetic fields: 500 G
(non-resonant, red); 313.82 G (resonant but $a_{\rm s}=a_{\rm bg}$, blue); 355
G (resonant, with $a_{\rm s} \neq a_{\rm bg}$, green), compared with
single-channel AQDT (black).  The coupled-channel calculations are truncated at
$L_{\rm max}=5$. Fully converged AQDT results are shown as black dotted lines
and are indistinguishable except at the highest energies. The length and energy
scaling factors $\bar{a}$ and $\bar{E}$ are 46~\AA\ and 218~$\mu$K for RbCs.
\label{fig:RbCs}}
\end{figure}

Figure \ref{fig:RbCs} compares calculations on RbCs at various magnetic fields
$B$, using the interaction potential of ref.\ \cite{Takekoshi:RbCs:2012}. In a
magnetic field $\sigma_\eta^{(1)}$ is no longer given by Eq.\
\ref{sig_eta_alt}, but it can still be simply calculated from S-matrix elements
\cite{Krems:FWQC:2004}. At $B=500$~G, the scattering non-resonant region and
the scattering length is close to its background value $a_{\rm s} = a_{\rm bg}
\approx 350\ {\rm \AA} \approx 7.5 \bar{a}$; $B=313.82$~G is in a region with
numerous overlapping resonances but where the scattering length is
coincidentally close to the background scattering length; and $B=355$~G is near
a resonance at a point where the scattering length is small, $a_{\rm
s}=12$~\AA. Full coupled-channel calculations in a magnetic field become
prohibitively expensive for large basis sets, so the coupled-channel results
are truncated at $L_{\rm max}=5$. AQDT results for a single channel with the
background scattering length are also shown in Fig.\ \ref{fig:RbCs}. In the
non-resonant case, AQDT again gives excellent results for both $\sigma_{\rm
el}$ and $\sigma_\eta^{(1)}$, similar to that seen for the single-channel case
with $a_{\rm s}=7.5\bar{a}$ in Fig.\ \ref{fig:LiYb}(d). In the resonant case
with the same scattering length, the results are again similar, except for a
resonant feature that in this case occurs near $2\bar{E}$; here
$\sigma_\eta^{(1)}$ shows a characteristic peak and trough because the
interference terms in Eq.\ \ref{sig_eta_alt} pass through both positive and
negative values as one of the phases sweeps through $\pi$. Even when the
scattering length is resonantly shifted from its background value, so that the
limitingly low-energy scattering is different, the cross sections rapidly
approach the ``universal'' form from the background channel once a few partial
waves contribute.

\section{Summary and Conclusions}

The cross section that controls thermalization of gas mixtures is the transport
cross section $\sigma_\eta^{(1)}$ and not the elastic cross section
$\sigma_{\rm el}$. We have investigated the behavior of both these cross
sections for the prototype systems LiYb and RbYb, which are of current
experimental interest. The two cross sections are identical in the pure s-wave
regime, but differ at higher energies, when additional partial waves contribute
to the scattering. Measurements at such energies are often desirable to
determine the {\em sign} of the scattering length as well as its magnitude. At
energies high enough for the sign to make a difference, $\sigma_\eta^{(1)}$ and
$\sigma_{\rm el}$ are significantly different. The differences can appear at
very low energies when the s-wave scattering length is close to $+2\bar{a}$,
since then there is a p-wave shape resonance close to threshold.

For more complex cases such as pairs of alkali-metal atoms, resonances may have
a large effect on s-wave scattering, but the cross sections nevertheless
approach the universal form based on the background scattering length once
several partial waves contribute to the scattering. In this regime the
distinction between $\sigma_{\rm el}$ and $\sigma_\eta^{(1)}$ is again
significant.

\acknowledgments

The authors are grateful to Daniel Brue for discussions about the alkali-Yb
systems and acknowledge support from the Engineering and Physical Sciences
Research Council under grant no.\ EP/I012044/1, and from EOARD under Grant
FA8655-10-1-3033.

\bibliography{../all,thermalizationRefs}

\end{document}